\documentclass[aps,twocolumn]{revtex4}
\usepackage{amsmath}
\usepackage{graphicx}

\begin{document}
\title{Shifting nodal-plane suppressions in high-order harmonic spectra from diatomic molecules in orthogonally polarized driving fields}
\author{T. Das and C. Figueira de Morisson Faria\\
Department of Physics and Astronomy, University College London,\\ Gower Street, London WC1E, 6BT, UK\\}

\begin{abstract}

We analyze the imprint of nodal planes in high-order harmonic spectra from aligned diatomic molecules in intense laser fields whose components exhibit orthogonal polarizations. We show that the typical suppression in the spectra associated to nodal planes is distorted, and that this distortion can be employed to map the electron's angle of return to its parent ion. This investigation is performed semi-analytically at the single-molecule response and single-active orbital level, using the strong-field approximation and the steepest descent method. We show that the velocity form of the dipole operator is superior to the length form in providing information about this distortion. However, both forms introduce artifacts that are absent in the actual momentum-space wavefunction. Furthermore, elliptically polarized fields lead to larger distortions in comparison to two-color orthogonally polarized fields. These features are investigated in detail for $\mathrm{O}_2$, whose highest occupied molecular orbital provides two orthogonal nodal planes. 
\end{abstract}

\date{\today}
\maketitle
\affiliation{$^1$Department of Physics and Astronomy, University College London, Gower Street, London WC1 6BT, UK}

\address{Department of Physics and Astronomy, University College London,\\ Gower Street, London WC1E 6BT, UK}

\section{Introduction}

Strong laser fields composed of waves with orthogonal polarizations are a useful resource for controlling strong-field phenomena, such as high-order harmonic generation (HHG) and its applications, for example in the creation of attosecond pulses \cite{Corkum_1994,Ivanov_1995,Antoine_1996,Chang_2004}.  They have also gained a great deal of attention as potential attosecond imaging tools. For instance, orthogonally polarized fields may be used to probe degenerate orbitals, molecules that are difficult to align, and also allow the reconstruction of molecular orbitals from a single-shot measurement \cite{Shafir_2009, Kitzler_2007, Kitzler_2005,Niikura_2010, Niikura_2011, Yun_2015}. 

This control is possible due to the physical mechanism behind HHG \cite{Corkum_1993}, namely the laser-induced recombination of an electron with a bound state of its parent molecule. Thereby, an electron is typically freed by tunnel ionization, acquires kinetic energy from the field while propagating in the continuum and, subsequently, releases this energy as high-frequency radiation if it recombines with its parent molecule.  This means that the HHG spectrum contains information about the atom or molecule from which it was generated, such as vibrational motion \cite{Baker_2006,Baker_2008,Li_2008}, electron motion \cite{Niikura_2005}  and electronic structure \cite{Itatani_2004,Worner_2009,Morishita_2008, Kitzler_2008}. This information can be used for example in the reconstruction of molecular orbitals \cite{Shafir_2009}. Furthermore, orthogonally polarized fields provide extra degrees of freedom. This introduces an angle of ionization and recombination for the electron, and allows one to steer the electron propagation in the continuum. For instance, depending on the field parameters, the electron may return at a particular angle, which can be controlled \cite{Kitzler_2005, Kitzler_2007,Niikura_2010} and varies with the harmonic frequency \cite{Kitzler_2008}.

Care must be taken, however, as there may be features that are not related to the target, but to the field itself. One option is to keep the wave along the minor polarization axis weak enough, so that these distortions are minimized \cite{Shafir_2009,Dudovich_2006}. 
On the other hand, a weak field limits the possibility of steering the electron trajectory. Another option is to understand how a non-vanishing field ellipticity modifies the information in the HHG spectra. This can then be used either to disentangle the influence of the field, and thus probe the molecular target, or to understand the electron dynamics in the continuum. 

For instance, it may happen that features that are purely structural may acquire dynamic aspects if the field ellipticity is non-vanishing. A good example are the interference patterns related to electron recollision in different centers in the molecule. For linearly polarized fields, these patterns are well understood and have been studied since the early 2000s \cite{Lein_2002,Lein_2002_2} (for reviews see, e.g., \cite{Lein_2007} and our recent publication \cite{Brad_2012}), at least within the single-active electron, single-active orbital approximation. In this case, the aligned diatomic molecule acts like the microscopic counterpart of a double-slit experiment, and the interference maxima and minima depend only on the internuclear distance and the molecule orientation with regard to the field. A particularly good method for assessing this type of interference is the strong-field approximation (SFA), which allows an intuitive interpretation of the problem in terms of electron orbits. For that reason, it has been widely employed in the study of molecules \cite{Kopold_1998,Madsen_2006,Madsen_2007,Chirila_2006,Faria_2007,Faria_2010,Brad_2011,Etches_2010,Odzak_2009,Odzak_2010}). The SFA, however, has serious limitations. Apart from the gauge dependence and its influence on the structural interference \cite{Chirila_2006,Faria_2007,Smirnova_2007}, different forms of recombination dipole matrix element affect this condition  \cite{Chirila_2007,Brad_2011,Granados_2012}. The most appropriate form to be used has raised considerable debate in the context of the tomographical reconstruction of molecular orbitals \cite{Zwan_2008,Hijano_2010,Zhu_2013}

If the field however is orthogonally polarized, the electron's angle of return is effectively incorporated in the two-center interference condition. This angle will depend on the orbit along which the electron comes back to its parent ion \cite{Das_2013}. Hence, different orbits will have different start and return times, which are heavily dependent on the field parameters and the harmonic energy. For coherent superpositions of orbits, this will cause a blurring in structural interference conditions. The outcome of many studies show such effects, but do not relate them to the electron's angle of return \cite{Odzak_2010,Odzak_2011}. In previous work, we have analyzed these effects in detail \cite{Das_2013}, and shown that they are present in the HHG macroscopic response for carefully chosen propagation conditions \cite{Das_2015}. This provides a tool for determining the electron's return angle in an experimental setting.  

Another well known structural feature is that, when a nodal plane is in alignment with the polarization of the field, there is a drop in HHG efficiency across the whole spectrum (see, e.g., \cite{Brad_2011,Smirnova_2009,McFarland_2008,Odzak_2009}). This suppression arises from the fact that nodal planes are areas of vanishing probability density in the wavefunction of a molecule. Vanishing probability density means that neither ionization nor recombination can take place \cite{Pavicic_2007,Abusamha_2009, Petretti_2010}. Further studies in \cite{Brad_2011b} compared the signals of nodal planes in isoelectronic homonuclear and heteronuclear diatomic molecules in HHG spectra. For the latter case the nodal planes were distorted into nodal surfaces. This caused the suppression in the spectrum to appear at different angles, in comparison to the homonuclear molecule. 

Similar distortions appear in the HHG spectra calculated in  \cite{Odzak_2010} for HHG in elliptically polarized fields, but they have not been analyzed. Therein, the suppressions related to nodal planes appear to shift and bend if the ellipticity of the driving field is increased \cite{Odzak_2010}. Our previous work \cite{Das_2013,Das_2015} indicates that the origin of these distortions lie on dynamic effects introduced by the electron's returning angle. This is an open question as previous publications that have addressed nodal-plane suppressions and elliptical fields \cite{Bhardwaj_2001,Lein_JPB_2003} have focused on ionization, but not recombination. They found that, although on their own nodal planes and elliptical field suppress ionization, the combination of both can in fact compensate for each other. This increases the HHG signal when the major polarization axis and the nodal plane are in alignment. 

In this paper we focus on the influence of the driving-field ellipticity on the HHG suppression caused by nodal planes. Using the equation for the effective shift of an returning electron presented in \cite{Das_2013},  we are able to predict where in the spectrum the nodal suppression will appear for a particular harmonic. Because this shift is orbit-dependent, the alignment angle for which the nodal-plane suppression appear will vary across the HHG spectrum. These features are investigated in detail for $\mathrm{O}_2$ in one- and two-color orthogonally polarized fields. In this work, we employ the single-active electron, single-active orbital approximation and neglect core dynamics. The latter issue has been addressed in, for instance, \cite{Smirnova_2009,Smirnova_2_2009}.

This article is organized as follows. In Sec.~\ref{Theory}, we provide the necessary theoretical background. This includes the generalization of the SFA to orthogonally polarized driving fields, and how the active orbital is modeled. We also revisit the orbit-dependent dynamic shift derived in \cite{Das_2013} (Sec.~\ref{Angle Of Return}).  
In Sec.~\ref{spectra}, we compute HHG spectra using one and two-color orthogonally polarized fields, and analyze the features encountered. This includes the nodal-plane distortions for individual orbits, the most convenient form of the strong-field approximation and the most favorable field configurations in order to observe the shifts. Finally, in Sec. \ref{conclusions}, we provide the main conclusions to be drawn from this work.  

\section{Model} 
\label{Theory} 
Throughout the paper, we will employ time-dependent fields composed of two orthogonal linearly polarized waves. We explicitly write the external electric field and the corresponding vector potential as
\begin{equation}
	\mathbf{E}(t)=E_{\parallel}(t)\hat{\epsilon}_{\parallel}+E_{\perp}(t)\hat{%
		\epsilon}_{\perp}
	\label{eq:Efield}
\end{equation}
and
\begin{equation}
	\mathbf{A}(t)=A_{\parallel}(t)\hat{\epsilon}_{\parallel}+A_{\perp}(t)\hat{%
		\epsilon}_{\perp},
	\label{eq:Afield}
\end{equation}
respectively, where the subscripts ($||$) and ($\perp$) designate field components parallel to the major and minor polarization axis, respectively.  The unit vector along the major and the minor polarization axis are denoted by $\hat{\epsilon}_{\parallel}$ and $\hat{\epsilon}_{\perp}$,
respectively. They are related to each other through $\mathbf{E}(t)=-d\mathbf{A}(t)/dt$. 
\subsection{Transition Amplitude}
\label{SFAamplitude} The SFA transition amplitude for HHG \cite%
{Lewenstein_1994} is given by
\begin{align}  \label{Tamp}
M(\Omega)&=-i \int^{\infty}_{-\infty} dt \int_{-\infty}^{t} dt^{\prime} \int d^{3}
\mathbf{p} \textbf{d}^*_{rec}(\mathbf{p}+\mathbf{A}(t)) \notag \\
&\times \textbf{d}_{ion}(\mathbf{p}+\mathbf{A}(t^{\prime}))e^{i S(t,t^{\prime },\Omega,\mathbf{p})} +c.c,
\end{align}
where the semi-classical action is given by
\begin{equation}  \label{action}
S(t,t^{\prime },\Omega,\mathbf{p})= -\frac{1}{2} \int^t_{t^{\prime }}[%
\mathbf{p}+\mathbf{A}(\tau)]^2 d\tau - I_p(t-t^{\prime}) + \Omega t
\end{equation}
and the ionization and recombination dipole matrix elements along the major polarization axis are 
\begin{equation}
d_{ion}(\mathbf{p}) =\langle \mathbf{p}|H_I(t^{\prime})|\Psi_0 \rangle
\label{dion}
\end{equation}
and 
\begin{equation}
d_{rec}(\mathbf{p+\mathbf{A}(t)})=\langle \mathbf{p+\mathbf{A}(t)}|\hat{\mathbf{d}}\cdot\hat{{\epsilon}}_{\parallel}| \Psi_0\rangle
\label{drec}
\end{equation}
respectively. The semi-classical action describes the propagation of an electron in the continuum between the time in which it ionizes, $t^{\prime }$, to the time $t$ when it recombines to its parent molecule. In the above-stated equations, $\hat{\mathbf{d}}$, $I_p$ and $\Omega$ give the dipole operator, the ionization potential, and the harmonic frequency, respectively.
The recombination prefactor (\ref{drec}) can be written in different forms, which will lead to different results. The form of the dipole operator should not be confused with the gauge \cite{Chirila_2007,Faria_2007,Brad_2011}, which determines how the Hamiltonian is written.  Explicitly, the length, velocity and acceleration forms of the dipole operator read $\hat{\textbf{d}}^{(l)}= \hat{\textbf{r}}$, $\hat{\textbf{d}}^{(v)}= \hat{\textbf{p}}$ and $\hat{\textbf{d}}^{(a)}=- \nabla V(\hat{\textbf{r}})$ respectively, where the hats denote operators. In this work we have used the length gauge, so that the interaction Hamiltonian in Eq.~(\ref{dion}) is given by  $H_I(t^{\prime})=\hat{\mathbf{r}}\cdot\mathbf{E}_{\parallel}(t^{\prime})$ and we consider the length and velocity forms of the dipole operator. 

 All the information about the structure of the molecule are contained within the ionization and recombination prefactors [Eq.~(\ref{dion}) and (\ref{drec})]. We neglect the motion of the nuclei and use the single active orbital approximation, which assumes that only the highest occupied molecular orbital (HOMO) contributes to the dynamics. We represent the HOMO by a linear combination of atomic orbitals (LCAO), which represents the HOMO wavefunction $\Psi _{0}(\mathbf{r})$ as follows,
\begin{equation}
\Psi _{0}(\mathbf{r})=\hspace*{-0.2cm}\sum_{a}c_{a }\hspace*{-0.1cm}\left[ \psi _{a }\hspace*{-0.1cm}\left( \mathbf{r}%
+\frac{\mathbf{R}}{2}\right) \hspace*{-0.1cm}+\hspace*{-0.1cm}(-1)^{\ell _{a }-m_{a }+\lambda
_{a }}\psi _{a }\hspace*{-0.1cm}\left( \mathbf{r}-\frac{\mathbf{R}}{2}\right) %
\right] \hspace*{-0.1cm},  \label{HOMOwf}
\end{equation}%
where $\psi _{a }(\mathbf{r})$, \textbf{R}, $c_a$ are the atomic orbitals, the internuclear distance, and the LCAO coefficients, respectively, while $\ell_{a} $ and $m_{a}$  refer to the orbital and to the
magnetic quantum number, respectively. The indices $\lambda_{a} =m_{a}$ correspond to gerade (g) and $\lambda_{a} =m_{a}+1$ to ungerade (u) orbital symmetry.

This means that we can write the dipole matrix element $d_{rec}(\mathbf{p}+\mathbf{A}(t))$ for
the wavefunction (\ref{HOMOwf}) in the length form as
\begin{eqnarray}
d^{(l)}_{rec}(\mathbf{p}(t))\hspace*{-0.1cm}&=\hspace*{-0.1cm}&\sum_a%
\hspace*{-0.15cm}c_a\left[ e^{i\mathbf{p}(t)\cdot \frac{\mathbf{R}}{2}%
}+(-1)^{\ell_a-m_a+\lambda_a}e^{-i\mathbf{p}(t)\cdot\frac{\mathbf{R}}{2}} \right]  \notag \\
&& \times i\partial_{p_{\parallel}(t)}\psi_{a}(\mathbf{p}(t)),\label{eq:dipolelenght}
\end{eqnarray}
and in the velocity form as 
\begin{eqnarray}
d^{(v)}_{rec}(\mathbf{p}(t))\hspace*{-0.1cm}&=\hspace*{-0.1cm}&\sum_a%
\hspace*{-0.15cm}c_a\left[ e^{i\mathbf{p}(t)\cdot \frac{\mathbf{R}}{2}%
}+(-1)^{\ell_a-m_a+\lambda_a}e^{-i\mathbf{p}(t)\cdot\frac{\mathbf{R}}{2}} \right]  \notag \\
&& \times p_{\parallel}(t)\psi_{a}(\mathbf{p}(t)),
\end{eqnarray}
where $\mathbf{p}(t)=\mathbf{p} + \mathbf{A}(t)$ and
\begin{equation}
\psi_{a}(\mathbf{p}(t))=\frac{1}{(2\pi)^{3/2}}\int d^3r \psi_{a}(%
\mathbf{r})\exp[-i \mathbf{r} \cdot \mathbf{p}(t)].
\end{equation}
In Eq.~(\ref{eq:dipolelenght}), the term related to the lack of orthogonality between bound state and continuum states that occurs in the SFA has been removed by hand. This is a widely used procedure, and it is related to the fact that this term blurs the two-center interference condition (see \cite{Chirila_2006,Smirnova_2007,Faria_2007} for discussions). 

 The reference frame of the molecule is rotated by the alignment angle $\theta_L$ with regard to the major polarization axis of the field. If we consider $xy$ as the polarization plane, this means that one may relate the $p_{\parallel}$, $p_{\perp}$ components to the components $p_x$, $p_y$ parallel and perpendicular to the molecular axis via
 
 \begin{equation}
 \left( \begin{array}{cc}
 p_{\parallel}& \\
 p_{\perp}& \end{array}\right) =
 \left(\begin{array}{ccc}
 \cos\theta_L & \sin\theta_L & \\
 -\sin\theta_L & \cos\theta_L & \end{array} \right)\left( \begin{array}{cc}
 p_{x}& \\
 p_{y}& \end{array}\right)
 \end{equation}
 
In the above-stated equation, we have taken into account that Gaussian-type orbitals have been employed in the construction of the HOMO, and that only $s$ and $p$ orbitals are included in the basis sets employed in this work. The orbitals used in this work have been computed with GAMESS-UK \cite{GAMESS}.  In this case, the derivative of the momentum-space wavefunction in the direction of the main polarization axis reads
\begin{align}
\partial_{p\parallel} \psi_a(\mathbf{p}) =& \left(-\frac{i}{2}\right)^{\ell_a} b_{a}c_{a}\pi^{\frac{3}{2}}\chi_a^{-\ell_{a}-\frac{3}{2}}e^{-(p_{\parallel}^2+p_{\perp}^2)/(4\chi_a)}\nonumber \\
& \times\left[{\mathcal K}_1(p_{\parallel}, p_{\perp}, \theta_L)+{\mathcal K}_2(p_{\parallel}, p_{\perp}, \theta_L)\right],\label{eq:Psiderp}
\end{align}
where
\begin{equation}
{\mathcal K}_1(p_{\parallel}, p_{\perp}, \theta_L)=-\ell_{a}\sin\theta_L(-p_{\parallel}\sin\theta_L+p_{\perp}\cos\theta_L)^{\ell_{a}-1}\label{eq:Psiderp1}
\end{equation}
and 
\begin{equation}
{\mathcal K}_2(p_{\parallel}, p_{\perp}, \theta_L)=-\frac{p_{\parallel}}{2\chi_a}(-p_{\parallel}\sin\theta_L+p_{\perp}\cos\theta_L)^{\ell_a}.\label{eq:Psiderp2}
\end{equation}
In Eqs.~(\ref{eq:Psiderp})--(\ref{eq:Psiderp2}),  $b_a$ and $\chi_a$ give the contraction and the exponential coefficients, respectively.

The transition amplitude (\ref{Tamp}) is  calculated using the saddle point
approximation, in which we solve Eq.~(\ref{Tamp}) by finding $t^{\prime}$, $t$ and $\mathbf{p}$ for which Eq.~(\ref{action}) is stationary. For the field in Eqs.~(\ref{eq:Efield}) and (\ref{eq:Afield}), we can re-write the action as 
\begin{align}
S(t,t^{\prime },\Omega ,\mathbf{p})& =-\frac{1}{2}\int_{t^{\prime
	}}^{t}d\tau \lbrack p_{||}+A_{||}(\tau )]^{2}  \notag  \label{actionEP} \\
	& -\frac{1}{2}\int_{t^{\prime }}^{t}d\tau \lbrack p_{\perp }+A_{\perp }(\tau
	)]^{2}-I_{p}(t-t^{\prime })+\Omega t. 
	\end{align}
This gives us the saddle-point equations
\begin{equation}
\frac{\partial S(t,t^{\prime },\mathbf{p})}{\partial t^{\prime }}=\frac{%
	[p_{||}+A_{||}(t^{\prime })]^{2}}{2}+\frac{[p_{\perp }+A_{\perp }(t^{\prime
	})]^{2}}{2}+I_{p}=0,  \label{eq:t'saddle}
\end{equation}%
\begin{equation}
\frac{\partial S(t,t^{\prime },\mathbf{p})}{\partial \mathbf{p}}%
=\int_{t^{\prime }}^{t}d\tau \lbrack \mathbf{p}_{||}+\mathbf{A}_{||}(\tau
)]+\int_{t^{\prime }}^{t}d\tau \lbrack \mathbf{p}_{\perp }+\mathbf{A}_{\perp
}(\tau )]=\mathbf{0},  \label{eq:psaddle}
\end{equation}%
and
\begin{equation}
\frac{\partial S(t,t^{\prime },\mathbf{p})}{\partial t}=\frac{%
	[p_{||}+A_{||}(t)]^{2}}{2}+\frac{[p_{\perp }+A_{\perp }(t)]^{2}}{2}%
+I_{p}-\Omega =0,  \label{eq:tsaddle}
\end{equation}%
respectively.

Physically, Eq.~(\ref{eq:t'saddle}) expresses the conservation of energy for the active electron upon tunnel ionization, for which there is no real solution. This reflects the fact that tunnel ionization is a quantum mechanical process. A return condition is imposed on the propagating electron by Eq.~(\ref{eq:psaddle}), fixing its intermediate momentum so that it returns to the site of its release, which is assumed to be the geometrical center of the diatomic molecule ($\mathbf{r}=0$). Lastly, Eq.~(\ref{eq:tsaddle}) gives the conservation of energy when the propagating electron recombines to it parent molecule. Here, the kinetic energy that the electron has acquired whilst in the continuum is converted in a high-harmonic photon of frequency $\Omega$. 

We employ the uniform approximation throughout when computing the transition probabilities associated with pairs of orbits. This method treats each pair collectively. When computing the transition probabilities associated with individual orbits we use the standard saddle-point approximation, which treats the orbits individually (for details see Ref.~\cite{Faria_2002}). This method can break down for one of the orbits when the imaginary part of the solutions diverges, leading to an increase in harmonic yield after the cutoff \cite{Faria_2002}. 

\subsection{Angle Of Return}
\label{Angle Of Return}
In \cite{Das_2013} we derived an expression for the effective shift, $\zeta (t,t^{\prime })$, which, when incorporated in the two-center interference condition, modifies the energy position of the interference minimum. The real part of this shift can be interpreted as an electron's angle of return with regard to the major polarization axis of the field. The shift is given by 
\begin{equation}
\zeta (t,t^{\prime })=\arctan \left[ \frac{p_{\perp }+A_{\perp }(t)}{p_{||}+A_{||}(t)}\right], \label{shift}
\end{equation}
where the stationary momentum can be obtained from Eq.~(\ref{eq:psaddle}) to give us 
\begin{equation}
p_{b }=\frac{-1}{t-t^{\prime }}\int_{t^{\prime }}^{t}A_{b
}(\tau )d\tau ,  \label{pstat}
\end{equation}
where $b=\parallel$ and $b=\perp$ refer to the momentum components along the major or minor polarization axis of the driving laser field.
From Eq.~(\ref{shift}) we can see that the angle with which the electron returns is dependent on the parallel and perpendicular field-dressed momentum of the returning electron. This implies that the shift will be strongly influenced by the vector-potential components at the electron's return time along each orbit. Furthermore, $ p_{\parallel }$ and $p_{\perp }$ are functions of the return and ionization times $t$ and $t^{\prime }$ according to the saddle-point Eq.~(\ref{pstat}). Hence, different orbits will have different angles of return. 

If an electron returns to the parent molecule with an angle with respect to the major polarization axis, we would expect that the position of the nodal-plane suppression in the HHG spectrum to be shifted to a different alignment angle. Hence, for orthogonally polarized fields the shift in the nodal-plane suppression is calculated using $\mathrm{Re}[\zeta(t,t^{\prime})]$.

\section{High-harmonic spectra}
\label{spectra}
In the results that follow, we use orthogonally polarized fields of the form
\begin{equation}
\mathbf{E}(t)=\frac{ E_{0}}{\sqrt{1+\xi ^{2}}}\left[\sin (\omega t)\hat{\epsilon}%
_{\parallel }+\xi \sin (n\omega t-2\pi \phi )\hat{\epsilon}%
_{\perp }\right] ,  \label{field}
\end{equation}%
where the frequency ratio of $n=1$ gives an elliptically polarized field and $n=2$ corresponds to an orthogonally polarized two color (OTC) field, for which  the frequency of the field component along the minor axis is twice that of the wave along the major axis. In Eq.~(\ref{field}), the strength of the field component along the minor polarization axis relative to its component along the major axis is determined by $\xi $, and the relative phase $\phi $ controls the time delay between both waves. The field has been normalized so that the overall time-averaged intensity $\langle \mathbf{E}^2(t) \rangle_t$ remains constant. This implies that  the total ponderomotive energy $U_p=\langle A_{\parallel}^2(t)\rangle_t/2+\langle A_{\perp}^2(t)\rangle_t/2$ is kept constant for elliptical fields ($n=1$), and that $U_p$ will decrease with $\xi$ for OTC fields ($n=2$) \footnote{In the bichromatic case, division of the field amplitude by an overall factor $\sqrt{1+\xi ^{2}/4}$ is required in order to keep $U_p$ constant. This factor, however, excludes an overall constant time-averaged intensity.}. 

Throughout we employ O$_2$ as a molecular target, which is particularly convenient since its HOMO is a $1\pi_g$ orbital. 
This leads to two nodal planes that are perpendicular to each other and produce suppressions in the HHG spectrum when the molecular axis of O$_2$ is aligned at $\theta_L$ = 0, $\pi/2$, $\pi$, and $3\pi/2$ with respect to the major polarization of the field. It is also possible to avoid the effects of two-center interference in the HHG spectrum by an appropriate choice of driving-field intensity. For clarity, in the results that follow we restrict the electron ionization times to the first half  cycle of the driving field. 

\subsection{Individual prefactors}
\label{NodalPlaneHHG}
\begin{figure}[tbp]
\noindent \hspace*{-0.5cm}
\includegraphics[scale=0.33]{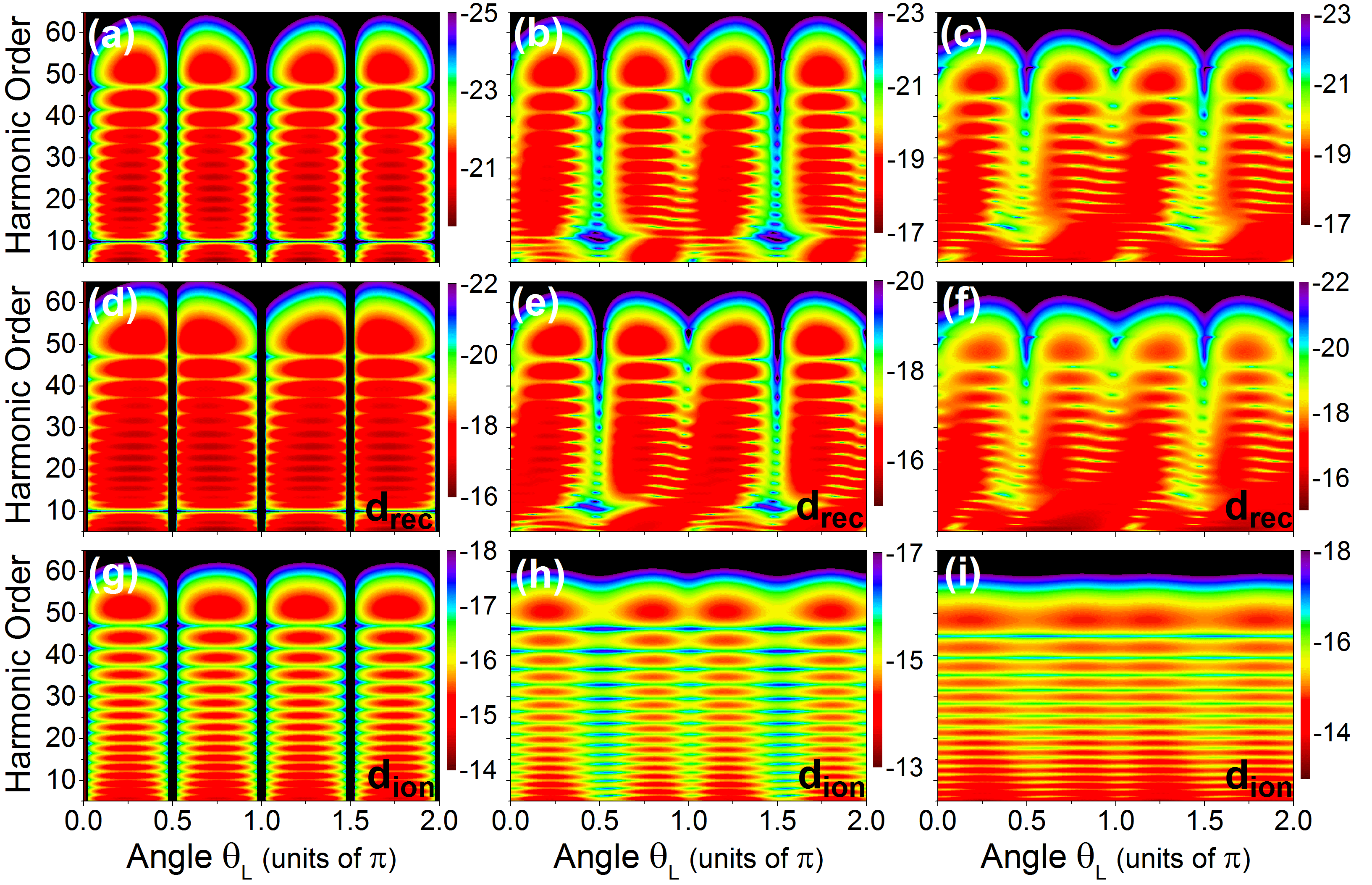}
\caption{(Color online) High-order harmonic spectra along the major polarization axis computed using the length form of the dipole operator for a coherent superposition of the dominant long and short orbits, as functions of the alignment angle $\protect\theta _{L}$ for O$_{2}$ ($I_p=$0.2446 a.u. and internuclear separation $R=2.28$ a.u.) in an elliptical field described in Eq.~(\protect\ref{field})  with $n=1$, $\protect\omega =0.057$ a.u., $I$=4$\times 10^{14}\mathrm{Wcm}^{-2}$ and time delay $\protect\phi =0.25$. The complete prefactor is calculated in the first row while only the recombination and ionization prefactors are used to calculate the spectrum in the second and third row respectively. The first, second and third column give an increasing value of the field ellipticity of $\protect\xi = 0,0.15$ and 0.3, respectively.}
\label{Fig1}
\end{figure}

We begin by focusing on how an elliptically polarized field modifies the position of the nodal-plane suppression in the HHG spectrum for the target molecule $\mathrm{O}_{2}$. In the first row of Fig.~\ref{Fig1} we display the transition probabilities $|M(\omega)|^2$ for a coherent superposition of the two dominant, shortest pair of orbits for increasing ellipticity. These orbits are well known in the literature as  the \textquotedblleft long orbit" and \textquotedblleft short orbit" \cite{Antoine_1996_PRL}, and correspond to electron excursion times of the order of three quarters of a field cycle. We find that the suppressions in the spectra begin to weaken and that the alignment angle for which they appear in the spectrum changes. This weakening and shifting in position decreases for increasing harmonic order and seems to behave differently for suppressions originally positioned at even and odd multiples of $\pi/2$ for linearly polarized fields. For $\theta_L=n\pi$, we observe more blurring and larger shifts, in comparison to the behavior near $\theta_L=(2n+1)\pi/2$.

In the remaining rows of Fig.~\ref{Fig1}, we show the contributions to the HHG spectrum from only the recombination or the ionization prefactor (second and third row, respectively). These figures show us that as the ellipticity of the field is increased the structure of the shifted nodal-plane suppressions is determined by the recombination prefactor. All the structure in the ionization prefactor is washed out for large enough ellipticity. This is in agreement with \cite{Lein_JPB_2003}, which found that, although the effect of the nodal plane and ellipticity of the field by themselves are detrimental to HHG, the combination of both can compensate for each other. 

\begin{figure}[tbp]
 \hspace*{-0.5cm}
 \includegraphics[scale=0.35]{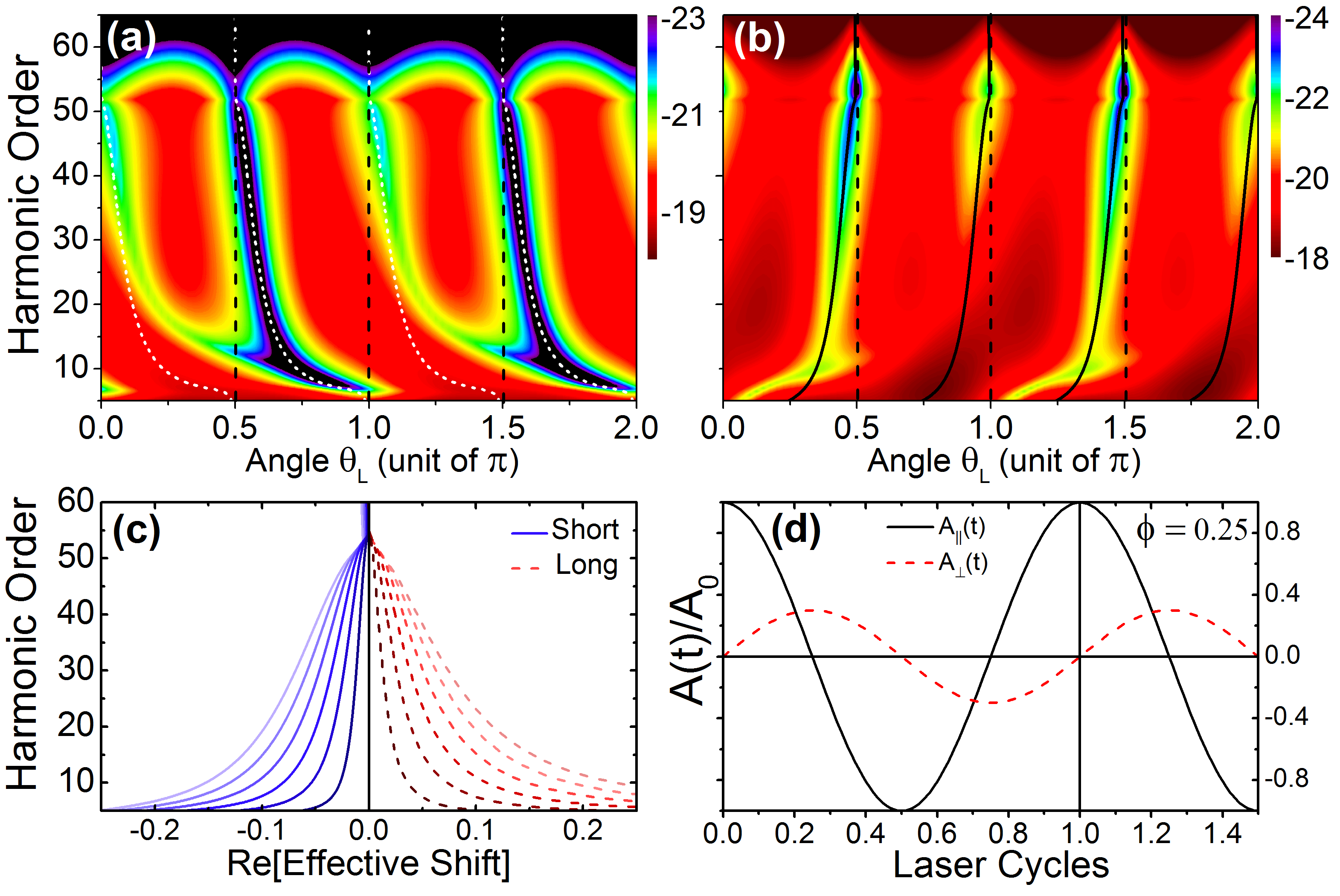}
\caption{(Color online) Panels (a) and (b) show the harmonic spectra along the major polarization axis as functions of the alignment angle $\protect\theta _{L}$ for O$_{2}$ in an elliptical field described in Eq.~(\protect\ref{field}), using the same parameters as in Fig.~\ref{Fig1} and the length form of the dipole operator. Panel (a) [Panel (b)] shows the individual contributions from the long [short] orbit. In panel (a), the shifted positions of the nodal-plane suppression calculated using $\mathrm{Re}[\zeta(t,t^{\prime})]$ [Eq.~(\protect\ref{shift})]  are indicated by the white short dashed curves, and in panel (b) they are given by the solid black lines. For comparison, we also indicate the position of the nodal-plane suppression for linearly polarized fields as the dashed black lines. The harmonic yield is given in a logarithmic scale. The increase in the harmonic yields after the cut-off observed in panel (b) is related to a breakdown of the standard saddle-point approximation for the short orbit (for details see Ref.~\protect\cite{Faria_2002}).
In panel (c) we have plotted the real parts of the effective shifts $\zeta(t,t^{\prime})$ as functions of the harmonic order computed for the long (red dashed curves) and short (blue solid curves) orbits in laser fields of increasing ellipticity and the same relative phase, intensity and frequency as in panels (a) and (b). The ellipticity has been increased from $\xi$ = 0 to  $\xi$ = 0.3 in increments of $\delta \xi$= 0.05. A lighter color indicates a higher ellipticity and a vanishing shift is indicated by a horizontal black line.
Panel (d) shows a schematic representation of the major and minor components of the vector potential \textbf{A}(t) for ellipticity $\xi$ = 0.3 and relative phase  $\phi$ = 0.25. The electron return time at  $t =2 \pi/\omega$ is indicated by the thick vertical black line in the figure. For simplicity, all fields have been normalized to the vector potential amplitude A$_0$ = E$_0/\omega$.}
\label{Fig2}
\end{figure}

\subsection{Individual Orbits and Different SFA Forms}
\label{individual_orbit}
In the upper row of Fig.~\ref{Fig2}, we show the transition probabilities $|M(\omega)|^2$ associated with individual orbits along which the active electron returns to the core, as functions of the alignment angle $\theta _{L}$. We consider the dominant, shortest pair of orbits. The contributions from the long and short orbits are displayed in panels (a) and (b), respectively. 

Throughout, we observe an excellent agreement between Eq.~(\ref{shift}) and the outcome of the SFA computations for the nodal plane-suppressions that are positioned at $\theta _L = (2n+1) \pi/2$ for linearly polarized fields. Furthermore, the positions of the suppressions are orbit dependent. This is expected, as  $\zeta(t,t^{\prime})$ depends on $t$ and $t^{\prime}$, which vary for the long and short orbits. In fact, the shift for the long orbit displaces the nodal-plane suppressions to the right, while for the short orbit this displacement is to the left. We also see that the displacement decreases for both orbits with increasingly higher harmonics. 
At the cutoff, the shifts vanish 
and the suppressions occur at $\theta _L = (2n+1) \pi/2$, as in the linearly polarized case. 

The above-stated observation
can be explained with  Fig.~\ref{Fig2}(c), in which the real parts of the effective shifts $\zeta (t,t^{\prime })$ are plotted for driving fields of increasing ellipticity. The case considered in the previous panels [Fig.~\ref{Fig2}(a) and (b)], i.e., $\xi=0.3$, is given by the outer curves. For the long orbit, this shift is positive. Hence, it will displace the suppression caused by the nodal plane towards larger alignment angles [see Fig.~\ref{Fig2}(a)]. A similar argument can relate the negative shift observed for the short orbit to the displacement to the left observed in Fig.~\ref{Fig2}(b).
  At and beyond the cut-off, the real parts of the shifts vanish. Consequently, the nodal-plane suppression will approach the position obtained for a linearly polarized field.
The reason for this is that around the cut-off the electron is expected to return at a crossing of the electric field, i.e., at a crest of the parallel vector potential [see Fig.~\ref{Fig2}(d)]. At such times, both orbits experience a vanishing perpendicular vector potential, which translates into a vanishing shift around the cut-off. Below the cut-off, the short and long orbits are subjected to equal but opposite perpendicular momenta, which increase for decreasing harmonic frequency. That is the reason why the nodal-plane suppressions are increasingly displaced in opposite directions for each orbit as we move to lower harmonics.

For the suppressions near even multiples of  $\theta_L = \pi/2$, the calculated effective shift does not fit the SFA outcome. The latter is strongly exaggerated for lower harmonics, and even meet the other shifted suppressions near the ionization threshold (see dot-dashed lines in the picture). We also see that the suppressions are more blurred than those encountered for $\theta_L=(2n+1)\pi/2$.   

\begin{figure}[tbp]
	\noindent \hspace*{-0.5cm}\includegraphics[scale=0.33]{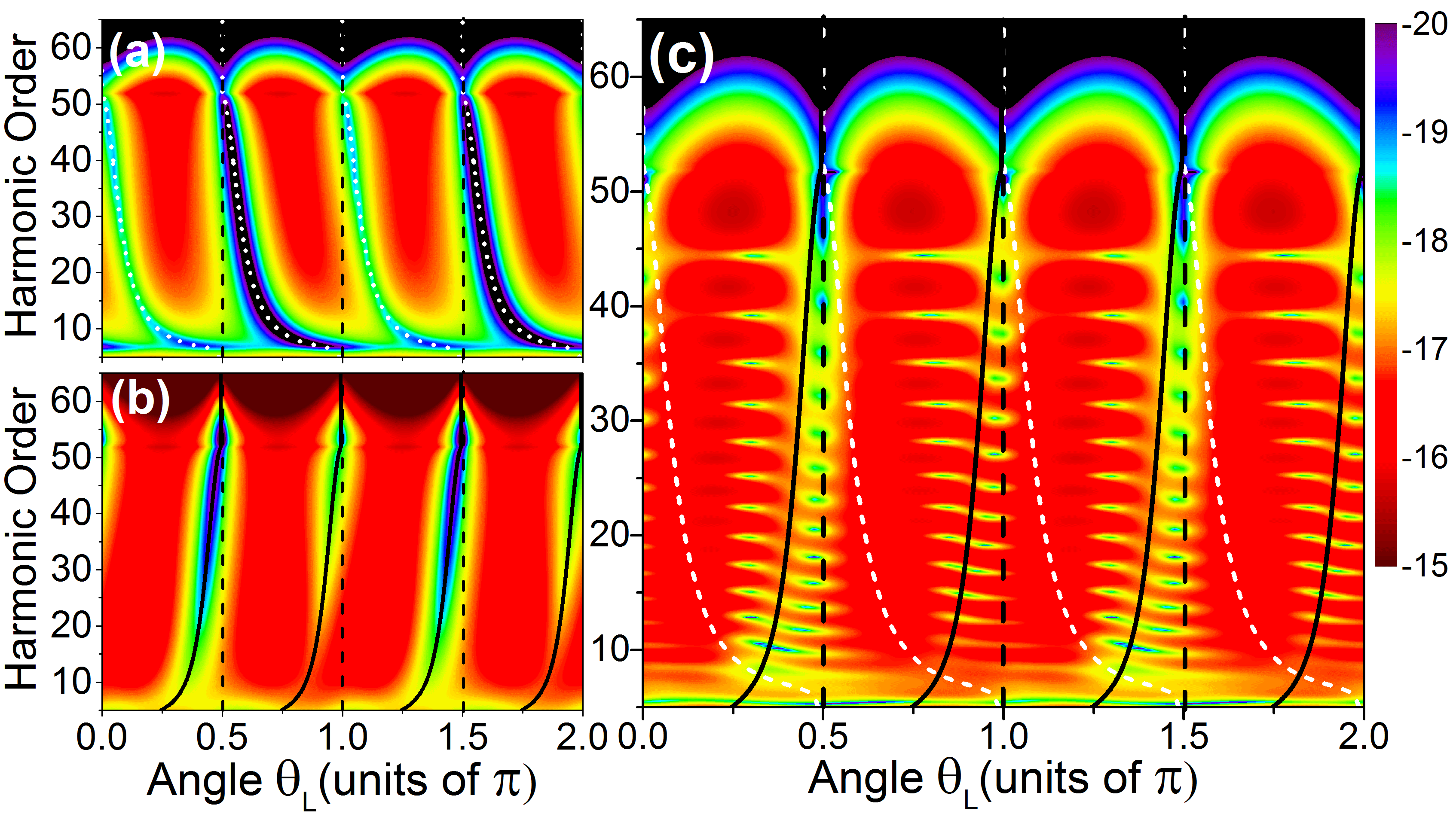}
	\caption{(Color online) Panels (a) and (b) show the harmonic spectrum for the long and short individual orbits respectively, along the major polarization axis as functions of the alignment angle $\protect\theta _{L}$ for O$_{2}$ in an elliptical field described in Eq.~(\protect\ref{field}) using the same parameters as in Fig.~\ref{Fig1}, but calculated using the velocity form of the dipole matrix elements. Panel (c) shows the harmonic spectrum for a coherent superposition of the dominant long and short orbits considered in Panels (a) and (b). The shifted positions of the nodal-plane suppression calculated using $\mathrm{Re}[\zeta(t,t^{\prime})]$ [Eq.~(\protect\ref{shift})]  are indicated by the white short dashed curves for the long orbit, and  by the solid black lines for the short orbit. For comparison, we also indicate the position of the nodal-plane suppression for linearly polarized fields as the dashed black lines.}
	\label{Fig4}
\end{figure}

If, instead, the matrix element $d^{(v)}_{rec}(\mathbf{p}\cdot \hat{\epsilon}_{\parallel})$ in the velocity form is used, the agreement between the SFA  and the analytical condition (\ref{shift}) improves significantly near $\theta_L =0, \pi$ and $2\pi$. There is, however, some blurring, if compared with the suppressions observed near odd multiples of $\pi/2$ for the low harmonic ranges. These results can be seen in Fig.~\ref{Fig4}, where we present the individual contributions of the long and short orbit [panels (a) and (b), respectively], together with the full spectrum [panel(c)], calculated using the velocity form of the SFA dipole matrix elements. 

These distortions are related to artifacts in the recombination dipole matrix elements, which leads to geometrical features that do not exist in the HOMO. In the present framework, the nodal structures are constructed in two ways. One may either employ nodes in the atomic orbitals at a \emph{single} center in the molecule, or the sum or subtraction of atomic orbitals at \emph{different} centers within the LCAO approximation. The former type of construction causes the suppressions at $\theta_L=n\pi$, while the latter lead to the suppressions at $\theta_L=(2n+1)\pi/2$. The velocity form of the SFA along the major polarization axis multiplies the momentum-space wavefunction by $p_{\parallel}$, while the length form of the SFA takes the partial derivative  $\partial_{p\parallel} \psi_a(\mathbf{p})$ of the atomic momentum wavefunctions used to construct the orbital [see Eq.~(\ref{eq:dipolelenght})]. Both procedures modify the nodal structures constructed using a single center. This spurious behavior becomes evident as the molecule rotates. 

\begin{figure}[tbp]
\noindent \hspace*{-0.6cm}\includegraphics[scale=0.43]{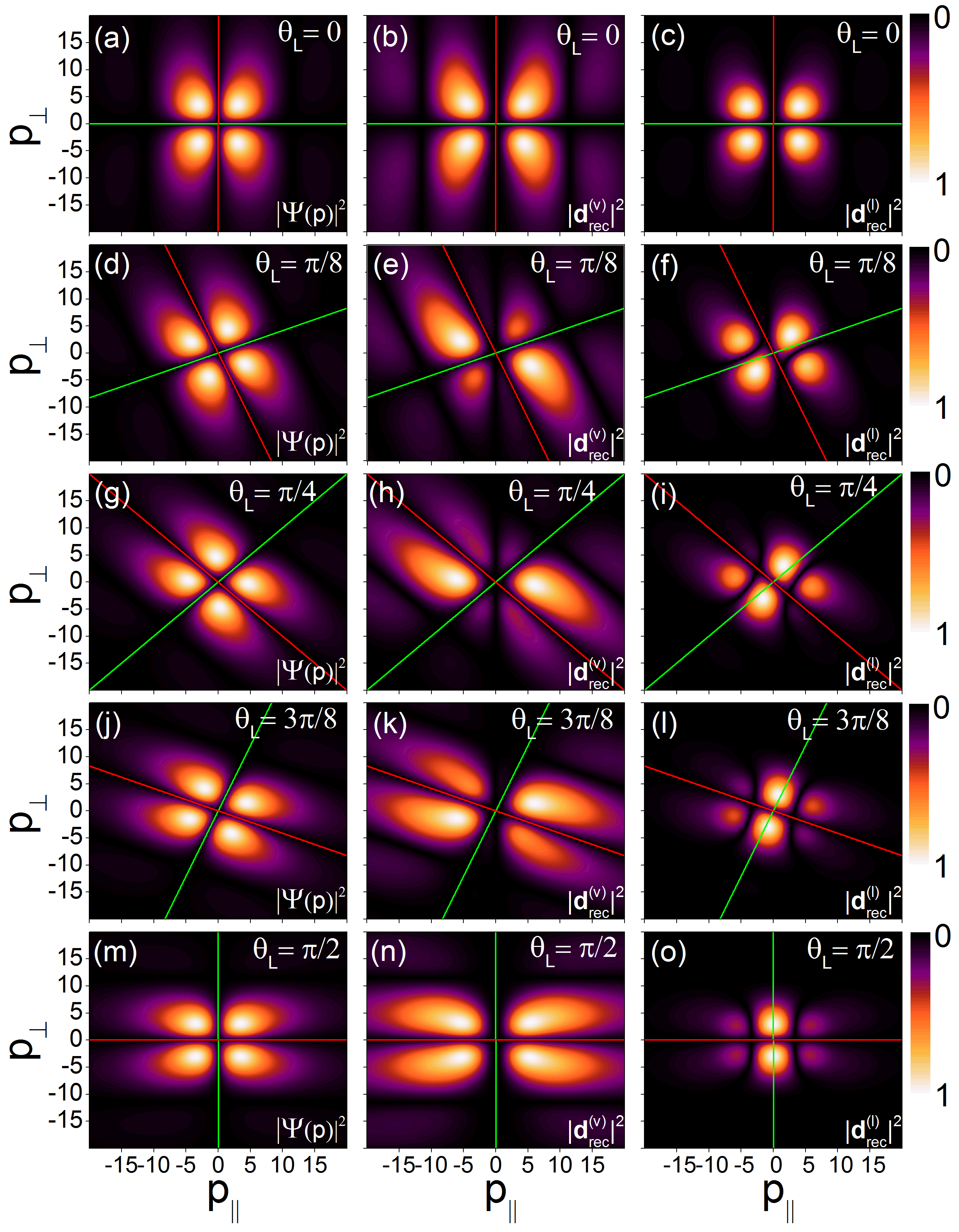}
\caption{(Color online) In the first, second and third columns we compare the probability density $|\Psi(\textbf{p})|^2$ in momentum space with the absolute squares of the dipole matrix elements $d^{(v)}_{rec}(\mathbf{p}\cdot \hat{\epsilon}_{\parallel})$ and $d^{(l)}_{rec}(\mathbf{p}\cdot \hat{\epsilon}_{\parallel})$ along the major polarization axis, respectively, for the HOMO of O$_2$.  The alignment angle $\theta_L$ is increased from the top row, $\theta_L =0$, to the bottom row $\theta_L = \pi/2$ in increments of $\delta \theta_L = \pi/8$. The HOMO of O$_2$ is a 1$\pi_g$ orbital where $I_p=$0.2446 a.u. and the internuclear separation is $R=2.28$ a.u.  The green and red lines in all the panels indicate the orientation of nodal planes constructed using atomic basis functions at single and different atomic centers, respectively. The quantity in each panel has been normalized by its maximum value.}

\label{Fig3}
\end{figure}
This is exemplified in Fig.~\ref{Fig3}, where we display the HOMO probability density $|\Psi(\mathbf{p})|^2$ for O$_2$ in momentum space, where $\Psi(\textbf{p})$ is the Fourier transform of Eq. (\ref{HOMOwf}), together with the absolute squares of the dipole matrix elements $d^{(v)}_{rec}(\mathbf{p}\cdot \hat{\epsilon}_{\parallel})$ and $d^{(l)}_{rec}(\mathbf{p}\cdot \hat{\epsilon}_{\parallel})$ along the major polarization axis, for several alignment angles $\theta_L$. For $\theta_L=0$, the three pictures are similar, with four lobes separated by two orthogonal nodal planes [see first row in the figure]. 
For $\theta_L\neq0$, however, this scenario changes, as shown in the remaining rows of the figure. While $|\Psi(\textbf{p})|^2$ does not alter its structure and merely rotates, for the velocity form there is an additional nodal plane at $p_{\parallel}=0$. For the length form, the behavior is more extreme and the node constructed with a single center begins to warp, split  and shift away from the original shape and orientation of $\psi(\textbf{p})$ indicated by the green lines. 

Both structures can be understood by inspecting the two prefactors. In the velocity form,  $p_{\parallel}\psi_{a}(\mathbf{p})$ implies that there will be a suppression at $p_{\parallel}=0$. For $\theta_L=n\pi$, this condition coincides with the nodal plane given by  $\tan \theta_L=p_{\perp}/p_{\parallel}$, which is obtained by imposing $\psi_{a}(\mathbf{p})=0$. For other angles, however, it leads to the spurious nodal structure. 
For the length form,  the two terms in  $\partial_{p\parallel}\psi_a(\textbf{p})$ lead, in general, to structures that are quadratic in $p_{\parallel}$. Specifically, Eq.~(\ref{eq:Psiderp1}) moves the suppression away from the axis $p_{\parallel}=0$ and the term given by Eq.~(\ref{eq:Psiderp2}) gives the above-mentioned warping. Once more, these spurious effects disappear for $\theta_L=n\pi$. 
These artifacts are overlooked by linearly polarized fields, as the electron's angle of return is always vanishing. This means that linearly polarized fields only probe the $p_{\parallel}$ axis in Fig. \ref{Fig3}. If the nodal planes are not parallel to this axis, the returning electron will ``see" a non-vanishing probability density and no suppression will occur. Hence, linearly polarized fields can only probe the nodal planes at multiples of $\pi/2$, for which the distortions cannot be seen along this axis.

The length or velocity form of the SFA dipole matrix elements has been a cause of much debate. 
 In the single-active electron, single-active orbital approximation, it is known that the velocity form is in superior in predicting structural interference minima, and provides the best agreement with the double-slit physical picture \cite{Chirila_2007}. However, for linearly polarized driving fields the spurious terms introduced by the length form are well understood and easy to eliminate. They are caused by the lack of orthogonality between the bound and continuum states that exists in the SFA \cite{Smirnova_2007,Faria_2007}, and are absent from the start in the expressions used in this paper  [see Eq.~(\ref{eq:dipolelenght})]. 

The results in Figs.~\ref{Fig2} to \ref{Fig3} tell us that, although for a linearly polarized field the form of the dipole operator may not make much difference, to the nodal suppressions in the HHG spectrum for an elliptically polarized field it does. This is because in this type of field the returning electron can probe dynamics of the wavefunction that would previously be unreachable.
This exposes other artifacts in both forms of the SFA dipole, which are more difficult to eliminate. Nonetheless, the velocity form provides better results, if compared to the length form.
 \subsection{Phase and Field Selection}
 \label{PhaseSelection}
\begin{figure}[tbp]
\hspace*{-0.5cm}
\noindent\includegraphics[scale=0.45]{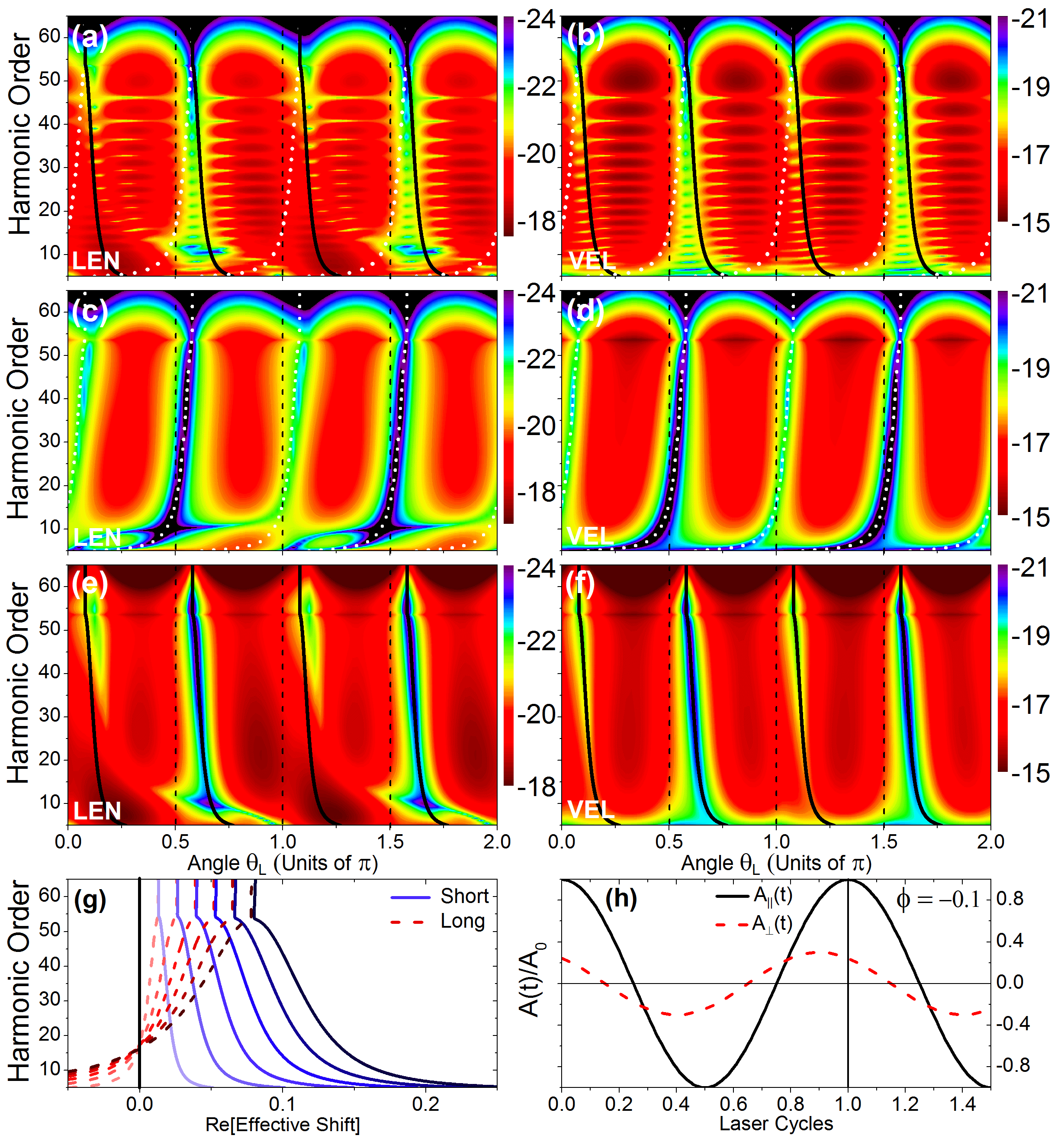}
\caption{(Color online) HHG spectra calculated  using the length (first column) and the velocity (second column) forms of the SFA. The first, second and third row give the coherent superposition of the dominant orbits [panels (a) and (b)], and the individual contributions of the long [panels (c) and (d)] and short orbits [panels (e) and (f)], respectively. 
The parameters used are the same as in Fig.~\ref{Fig1}, but with a time delay $\phi=-0.1$ between the parallel and perpendicular waves. 
The black dashed lines indicate the position of the nodal-plane suppressions for a linearly polarized field, whilst the white short dashed and solid black curves give the calculated position of the suppression for the long and short orbit, respectively, for elliptically polarized fields. 
In panel (g) we have plotted the real parts of the effective shifts $\zeta(t,t^{\prime})$ as functions of the harmonic order computed for the long (red solid curves) and short (blue dashed curves) orbits in laser fields of increasing ellipticity and the same relative phase, intensity and frequency as in panels (a) to (f). The ellipticity is increased from $\xi$ = 0 to  $\xi$ = 0.3 in increments of $\delta \xi$= 0.05. A lighter color indicates a higher ellipticity and a vanishing shift is indicated by a vertical black line.
Panel (h) provides a schematic representation of the major and minor components of the vector potential A(t) for ellipticity  $\xi$ = 0.3 and relative phase  $\phi$ = -0.1. The electron return time at  $ t =2 \pi/\omega$ is indicated by the thick vertical black line in the figure. For simplicity, all
fields have been normalized to the vector potential amplitude A$_0$ = E$_0/\omega$.}
\label{Fig5}
\end{figure}

In Figs.~\ref{Fig2} and \ref{Fig4} the shift of the nodal-plane suppression is quite large, especially for lower harmonics. However, around the cutoff the shift vanishes.
Since, however, the long and the short orbit merge at the cutoff, this would be the best region to observe the shift if a coherent superposition of orbits is taken into consideration \cite{Das_2013}.  Furthermore, if one wishes to observe these shifts experimentally, it would be more convenient if they occurred for the whole harmonic range in the spectrum. In this section we will discuss field choices which are favorable to this behavior.

Since the shifts are strongly dependent on the time delay between the parallel and perpendicular waves, choosing a different relative phase $\phi$ changes the behavior of the shift. An example is provided in Fig.~\ref{Fig5}, for which the parallel and perpendicular driving waves have a phase difference of $\phi=-0.1$. 
In Fig.~\ref{Fig5}, we show the SFA transition probabilities computed for this phase, using the whole dominant pair, the long and the short orbit [first, second and third row, respectively]. For comparison, we include results in the length [panels (a), (c) and (e)] and velocity [panels (b), (d) and (f)] forms of the SFA. 
 In contrast to what has been observed in the previous figures, the shifts are now present for the whole harmonic range, including the cutoff region.  Our results indicate that the shift would be easier to observe for the short orbit, as it is much larger in this case. This is convenient for experiments as the short orbit is much easier to isolate through phase matching \cite{Das_2015}. 
 Fig.~\ref{Fig5}(g), in which $\mathrm{Re}[\zeta (t,t^{\prime})]$ is plotted, is also markedly different from  Fig.~\ref{Fig2}(c). For instance, for the long and short orbit, the residual shift at the cut-off is positive. This is due to the fact that $A_{\perp}(t)$ is non-vanishing and positive at the cutoff return times [see Fig.~\ref{Fig5}(h)]. Once more, we see a better overall agreement between the analytical condition and the velocity form of the SFA.  For the coherent superposition of the two orbits [Figs.~\ref{Fig5}(a) and (b)], despite some blurring in the plateau and threshold harmonics, the shift can be seen very clearly at the cutoff. For all nodes, only the velocity form gives the correct shifts [Fig.~\ref{Fig5}(b)]. 

\begin{figure}[tbp]
\noindent
\includegraphics[scale=0.33]{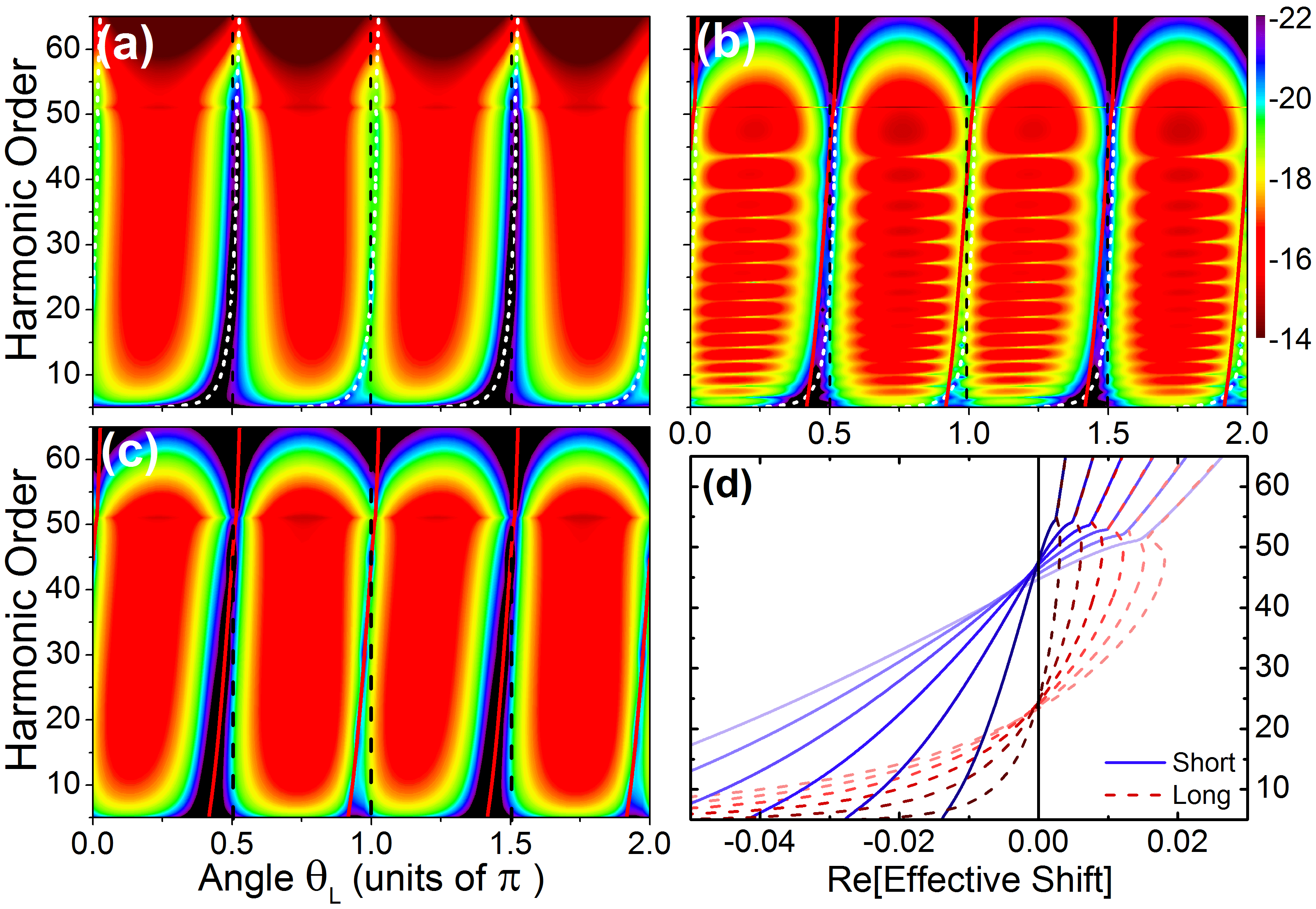}
\caption{ (Color online) HHG spectra along the major polarization axis as functions of the alignment angle $\protect\theta _{L}$ for O$_{2}$ (ionization potential $I_p=$0.2446 a.u. and internuclear separation $R=2.28$ a.u.). The parameters used are the same as in Fig.~\ref{Fig1}, but with $n=2$ for the perpendicular wave, which is in phase ($\phi=0$) with the parallel component of the laser field. Panels (a) and (c) give the individual contributions to the HHG spectrum from the long and short orbit respectively, whilst (b) shows the coherent superposition of these orbits. The black dashed lines indicate the positions of the nodal-plane suppressions in the spectrum for a linearly polarized field, whilst the white short dashed and red curves give the calculated position of the suppression for the long and short orbit, respectively, for  elliptically polarized fields. The harmonic yield is given in a logarithmic scale. The increase in the harmonic yields after the cutoff observed in panel (a) is related to a breakdown of the standard saddle-point approximation which occurs to the long orbits for this particular phase difference (for details see Ref.~\protect\cite{Faria_2002}). In panel (d) we have plotted the real parts of the effective shifts $\zeta(t,t^{\prime})$ as functions of the harmonic order computed for the long (red dashed  curves) and short (solid blue curves) orbits in laser fields of increasing ellipticity and the same relative phase, intensity and frequency as in panels (a),(b) and (c). The ellipticity is been increased from $\xi$ = 0 to  $\xi$ = 0.3 in increments of $\delta \xi$= 0.05. A lighter color indicates a higher ellipticity and a vanishing shift is indicated by a horizontal black line. }
\label{Fig6}
\end{figure}

One should bear in mind, however, that for elliptically polarized fields the recollision probability of the electron decreases as $\xi$ increases, which may lead to low HHG efficiency. For two-color orthogonal fields, on the other hand, the electron has higher probability of returning to the parent molecule as the strength of the perpendicular field is increased. For this reason, it is useful to compare computations using these two types of field.

In Fig.~\ref{Fig6}, we present HHG spectra computed for two-color fields. We consider a coherent superposition of the two dominant orbits [panel(b)], together with the individual contributions $|M(\omega)|^2$ from the long and short orbits [panels (a) and (c), respectively]. We have taken the relative phase $\phi=0$, which produces fairly large shifts for two-color fields. The figure shows that the shifts in the nodal-plane suppressions are much smaller than those obtained in the elliptically polarized case. For instance, for $\xi$ = 0.3,  an elliptical field may leads to shifts up to Re[$\zeta(t,t^{\prime})$] =0.25, while for a two-color field  the effective shift reaches up to Re[$\zeta(t,t^{\prime})$] = 0.05. This can be seen by comparing  the effective shifts in Fig.~\ref{Fig6}(d) with the elliptical-field examples in Figs.~\ref{Fig2}(c) and \ref{Fig5}(g). Hence, a two-color field would be less suitable for finding the shift in an experimental setting, despite the higher probability of return. 
 
\section{Conclusions}
\label{conclusions}

In this paper, we study high-order harmonic generation in aligned diatomic molecules in elliptical and  two-color orthogonally polarized laser fields. Our results show that, using fields of non-vanishing ellipticity, one may infer the angle with which an electron returns to its parent ion from HHG spectra, by relating it to distortions in the nodal-plane suppressions. While for linearly polarized fields these suppressions are well known and occur at fixed alignment angles throughout the spectra, if the fields are orthogonally polarized they are orbit- and harmonic-dependent. They can be controlled by changing the driving-field parameters, such as the relative phase, intensity and frequency ratios between the two orthogonal waves. We provide an analytic expression for this shift, which is checked against the strong-field approximation using the steepest descent method. As a testing ground, we have employed the HOMO of O$_2$, which exhibits two orthogonal nodal planes and no two-center interference minimum for the parameter range of interest, in our computations. 

Within our model, the suppressions in the spectra are caused by the recombination dipole matrix element in the SFA. The analytic condition works well for individual orbits if the nodal planes are constructed using atomic basis functions at different centers, but discrepancies arise for nodal planes in which basis functions at single centers are used. These discrepancies expose limitations in the recombination dipole matrix elements, which are overlooked for linearly polarized fields.  

In fact, we show that for specific alignment angles $\theta_L=n\pi/2$ the suppressions in these matrix elements coincide with the nodal planes. However, this is not the complete picture and spurious structures arise for other angles. These effects occur both for the length and the velocity form of the dipole operator. However, while for the velocity form they lead to a light blurring around the analytical condition, in the length form they are very extreme and lead to exaggerated distortions. Hence, the length form of the SFA should be avoided when mapping nodal planes using orthogonal fields. 

 One should note, however, that, because the shifts are orbit dependent, they may be difficult to extract unless either the cutoff region or a particular return event can be singled out. For coherent superpositions of orbits these patterns are blurred as there are many possible return angles. This is a similar situation to that encountered in our previous publication \cite{Das_2013}, in which the angle of return was incorporated in the structural, two-center interference condition.  This problem can however be solved through propagation, as the two dominant orbits phase match differently  \cite{Balcou_1997,Gaarde_2008}. This allows a high degree of control on which orbit is dominant in which spatial region \cite{Zair_2008,Auguste_2009}. 
 
 In \cite{Das_2015} we proposed a method to single out shifts for one trajectory in and experimental setting. This involved using polarization gating \cite{Hoffmann_2014} and phase-matching conditions to remove the contributions to the HHG spectrum of one of the dominant trajectories. In particular the short orbit is very convenient for observing these effects, as it phase matches on axis, for which the HHG signal is strong.  
 The same could be applied to the nodal-plane suppressions, for which there are two main advantages with regard to the shifted two-center interference condition. First, the distortions caused by the angle of return near nodal planes are in principle easier to identify as the suppressions are stronger. Second, with the right choice of phase difference, this suppression can be shifted across the spectra. 
 
 We have also found that elliptically polarized fields provide better conditions for observing the shifts experimentally, in comparison with OTC fields. First, using an elliptically polarized field gives rise to a much larger shift. Second, in \cite{Das_2013} we found that using a two-color orthogonal field caused the shift to flip sign every half cycle. This was because the parallel component $A_{\parallel}(t)$ of the vector potential would change sign every half cycle, but the perpendicular component $A_{\perp}(t)$ would not, causing $\zeta(t,t^{\prime})$ to change sign. In \cite{Das_2015}, we have avoided this problem by using a few-cycle pulse in which a specific cycle was dominant in the region of interest. For elliptical fields, however,  there is no need to restrict ionization events to a single half cycle. Hence, any pulse length can be employed. Thus, the present work proposes a way to extract an electron's angle of return using nodal planes as tools, which is valid as long as the single-active electron and orbital approximation holds. This seems to be the case for the plateau, as there is evidence that structural effects are dominant in this region \cite{Kato_2011,Rupenyan_2013}. 

\section*{Acknowledgments}
 This work has been funded by the UK EPSRC (grant EP/J019240/1 and doctoral training prize) and by UCL (Impact Studentship). We thank M. Kitzler and B.B. Augstein for useful discussions.

\end{document}